\documentclass{icrc2009}

\usepackage{graphicx}   
\usepackage[caption=false]{caption}    
\usepackage[font=footnotesize]{subfig} 
\usepackage{fixltx2e}
\usepackage{url}
\usepackage{amssymb}
\usepackage{multirow}

\newcommand{\shorttitle}[1]%
{\markboth{Proceedings of the 31\MakeLowercase{$^{st}$} ICRC, {\L}\'{o}d\'{z} 2009}{#1} }
\newcommand{\etal}{\MakeLowercase{\textit{et al. }}} 

\begin{document}
\title{Results of LUNASKA lunar Cherenkov observations at the ATCA}

\author{\IEEEauthorblockN{Clancy James\IEEEauthorrefmark{1,2},
			  Ron Ekers\IEEEauthorrefmark{3},
                          Chris Phillips\IEEEauthorrefmark{3},
                           Ray Protheroe\IEEEauthorrefmark{1},\\
                           Paul Roberts\IEEEauthorrefmark{3},
			   Rebecca Robinson\IEEEauthorrefmark{4},
			   Jaime Alvarez-Mu\~niz\IEEEauthorrefmark{5},
			   Justin Bray\IEEEauthorrefmark{1}}
                            \\
\IEEEauthorblockA{\IEEEauthorrefmark{1}School of Chemistry \& Physics, Univ.\ of Adelaide, SA 5005, Australia}
\IEEEauthorblockA{\IEEEauthorrefmark{2} Department of Astrophysics, Radboud University Nijmegen, Postbus 9010, 6500 GL Nijmegen,
The Netherlands}
\IEEEauthorblockA{\IEEEauthorrefmark{3}Australia Telescope National Facility, Epping, NSW 1710, Australia}
\IEEEauthorblockA{\IEEEauthorrefmark{4}School of Physics, Univ.\ of Melbourne, VIC 3010, Australia}
\IEEEauthorblockA{\IEEEauthorrefmark{5}Depto. de F\'\i sica de Part\'\i culas \& Instituto Galego de F\'\i sica de Altas Enerx\'\i as, \\ Univ.\ de Santiago de Compostela, 15782 Santiago de Compostela, Spain }
}
\shorttitle{C.W.~James \etal LUNASKA observations -- results.}
\maketitle

\begin{abstract}
We present the results of our (the LUNASKA project's) initial search for pulses of coherent radio Cherenkov radiation from UHE neutrino and cosmic ray interactions in the outer layers of the Moon. Three of the six $22$~m antennas of the Australia Telescope Compact Array (ATCA) were implemented with our purpose-built pulse detection hardware and used to observe the Moon for two three-night periods during 2008. The observation dates and beam pointing position were chosen to maximise sensitivity to Centaurus A and the Galactic Centre (nominally Sgr A) region. We present the results of our (now complete) data analysis of order $10^7$ candidate events -- though we find no evidence for lunar Cherenkov pulses, our use of a wide (600~MHz) bandwidth has made our observations the most sensitive yet using the lunar technique.
\end{abstract}

 
\section{Introduction}

The lunar Cherenkov technique \cite{Dagkesamanskii} is a method to use radio-telescopes to detect ultra-high energy cosmic rays (CR) and neutrinos ($\nu$). By observing the short-duration ($\sim$few nanosecond) pulses of coherent Cherenkov radiation emitted from particle cascades via the Askaryan Effect \cite{Askaryan} in the Moon's outer layers (nominally the regolith), the primary particles initiating the cascades may be identified. Our collaboration (LUNASKA) aims to develop the technique to be used with the next generation of giant radio-arrays. Here, we present the results of our two preliminary UHE particle searches using this technique with three antennas at the Australia Telescope Compact Array (ATCA) during February and May 2008.

We leave to another contribution (Ref.\ \cite{experimental}) the details of our specialised equipment, and here concentrate on the results of our search for lunar Cherenkov pulses. We first describe the observations, in which we targeted Centaurus A and the Galactic Centre region, then detail the procedures used to discriminate between random noise events, RFI, and lunar-origin pulses. We calculate an effective aperture for our experiment to UHE $\nu$, and since our pulse search could identify no lunar pulses, this aperture is used to set an upper limit on an isotropic flux of UHE neutrinos. Finally, we calculate our exposure as a function of UHE $\nu$ arrival direction, and compare this exposure with the exposure to Cen A and Sgr A from other experiments. Since we believe that current simulation techniques are inappropriate to modelling UHE CR hitting the Moon, we leave a calculation of our aperture/limit to UHE CR to a later work \cite{JP}.

\section{Observations}

We implemented three of the six $22$~m ATCA antennas with our specialised pulse detection and dedispersion hardware, and observed over a continuous $600$~MHz band in dual linear polarisations between $1.2$ and $1.8$~GHz. This apparatus is described in greater detail in our other contribution (Ref.\ \cite{experimental}) -- briefly, the equipment dedispersed, sampled ($2.048$~GHz, $8$-bit), and scanned the data streams from both polarisations at each antenna for candidate pulses, and returned a data buffer of length $256$~samples upon detecting a signal greater than an adjustable threshold. A clock time accurate to $0.5$~ns was also returned with each trigger. Each antenna operated independently due to data transmission constraints -- since this limited our sensitivity, we set each antenna to trigger at approximately $100$~Hz ($5$\% dead-time) to partially compensate.

A summary of our main observing runs and observation time in February and May 2008 is given in Table \ref{obstbl}. The February  run was tailored to `target' a broad ($\gtrsim 20^{\circ}$) region of the sky near the Galactic Centre, harbouring the closest supermassive black hole to Earth and potential accelerator of UHE CR (or site of decaying super-heavy dark matter). For these runs we pointed the antennas towards the lunar centre, since this mode maximised coverage of the entire lunar limb, from which we expect to see the majority of pulses. In this mode, we achieved the greatest total effective aperture.

Our May observing period targeted Centaurus A only, a nearby active galaxy which could potentially account for multiple UHE CR events as observed by the Pierre Auger observatory \cite{AugerScience07}. Regardless of their source, this suggests the likelihood of an accompanying excess of UHE neutrinos, and we do not exclude the possibility of seeing the UHE CR themselves. We therefore pointed the antennas at that part of the lunar limb closest to Cen A in order to maximise sensitivity to UHE particles from this region, as described in Ref.\ \cite{JP}.

\begin{table}
\begin{center}
\begin{tabular}{l| c c c | c c c}
\hline
 & \multicolumn{3}{c||}{Feb '08} & \multicolumn{3}{c|}{May '08} \\
\hline
\hline
Day	&	$26^{th}$ & $27^{th}$ & $28^{th}$ & $17^{th}$ & $18^{th}$ & $19^{th}$ \\

$t_{\rm obs}$ (mins) & 239	& 319	& 314	& 324	& 376	& 440 \\
$t_{\rm eff}$ (mins) & 204	& 277	& 274	& 224	& 274	& 316 \\
\hline
\end{tabular}
\caption{Observation dates (nights thereof), total time $t_{\rm obs}$, and effective observing time $t_{\rm eff}$ after subtracting trigger dead-time, spent observing the Moon in detection mode of our observing runs.}
\label{obstbl}
\end{center}
\end{table}

For each period, we observed nominally between the hours of $10$~pm and $6$~am local time. In 2008 we had to align the clocks manually by observing the bright quasar 3C273 and correlating the emission between antennas, with a $0.25$~ns timing uncertainty. Since $T_{\rm sys}$ measurements were available over only part of our bandwidth, we used the Moon as our absolute sensitivity calibrator, with a $2$\% error.

\section{Data and Pulse search}

\begin{figure}[!t]
 \centering
 \includegraphics[width=2.5in]{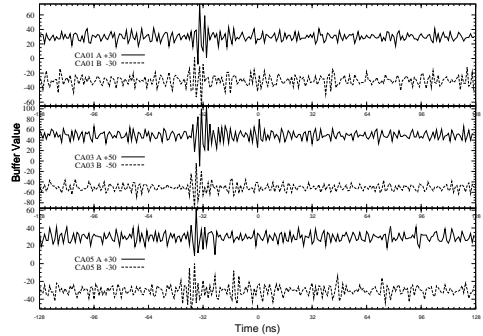}
 \caption{Typical short-duration RFI event.}
 \label{fig01}
\end{figure}

\begin{figure}[!t]
 \centering
 \includegraphics[width=2.5in]{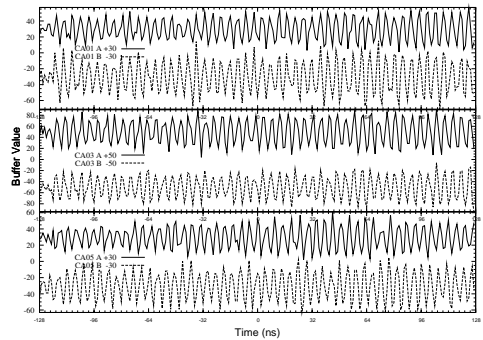}
 \caption{Typical long-duration RFI event.}
 \label{fig02}
\end{figure}

Our data set consisted of a few million triggers per antenna generated while pointing at the Moon. We therefore wrote a simple algorithm to reject at least $99.99$\% of the random coincidences and RFI events, so the remaining data could then feasibly be inspected `by-eye'. For this we used the timing criteria between antennas, since our three-antenna array allowed us to identify the arrival direction with extremely good accuracy.  A preliminary search first identified all threefold coincidences -- triggers on all three antennas within a time-envelope corresponding to the light travel-time between them -- using the raw clock times at each antenna. Because of the large number of RFI events, this left of order $100,000$ such coincidences. The relative clock times were then adjusted by pair-wise  cross-correlations of the buffers, to allow for different antennas triggering on different parts of the signal. Events with a significant time-structure -- e.g.\ Fig.\ \ref{fig01} -- produced a single strong correlation peak, with three pair-wise time adjustments that closed to zero. Those dominated by narrow-band RFI -- e.g.\ Fig.\ \ref{fig02} -- tended to produce many peaks in the correlation. While this meant picking the right correlation peak was difficult, and usually the resulting adjustments in the offsets did not close to zero, this `problem' actually gave an easy method of eliminating such events.

Using the adjusted time offsets and the known position of the Moon, a search over all remaining events produced $60$ coincidences where each pair-wise time offset was consistent with the lunar position, of which only $44$ resembled Fig.\ \ref{fig02} and could be discarded. For our baselines, the Moon was usually resolved by $O\sim30$ `slices' perpendicular to the baseline, and of the remaining $16$ candidates, none produced three time offsets consistent with the same location on the Moon, and only one was of sufficiently short duration to resemble a lunar Cherenkov pulse. Thus we concluded that no identifiable lunar Cherenkov pulses were observed.

\section{Effective Experimental Aperture}

\begin{figure}[!htb]
 \centering
 \includegraphics[width=2.5in]{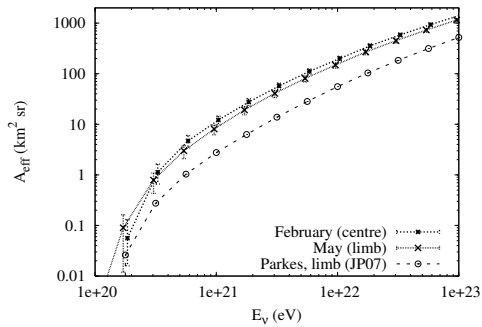}
 \caption{Isotropic apertures to UHE $\nu$, compared to the result for Parkes from \cite{JP}, assuming the existence of a sub-regolith layer.}
 \label{iso_app_atca}
\end{figure}

The instantaneous aperture, and consequent limits, from the experiment were calculated as per James and Protheroe \cite{JP}. In all cases, we assume the existence of an effectively-infinite sub-regolith layer, and ignore secondary $\mu$ and $\tau$ from the charged-current interactions of $\nu_{\mu}$ and $\nu_{\tau}$. Since both the trigger threshold and difference between assumed and actual STEC changed with time, so did the detection probability. We included the effects of finite sampling and approximate dedispersion by multiplying the simulated signal by a factor of order $0.9$, calculated as per Fig.\ 4 in Ref.\ \cite{experimental}. Therefore, we calculated maximum and minimum apertures for each pointing configuration, and an approximate mean effective aperture, using the median values for the trigger thresholds. These are shown in Fig.\ \ref{iso_app_atca}. The detection threshold of a few$\times 10^{20}$~eV is comparable to that of the Parkes experiment, since the increased bandwidth compensated for the reduced collecting area. Also, it is lower for the May observations due to the lower $T_{\rm sys}$ and greater sensitivity to the limb. However, at high energies, the ATCA observations (esp.\ for the centre-pointing configuration of February) see much more of the limb, and so are more sensitive than the Parkes experiment.

\begin{figure}[!t]
 \centering
\includegraphics[clip=true,width=2.5in]{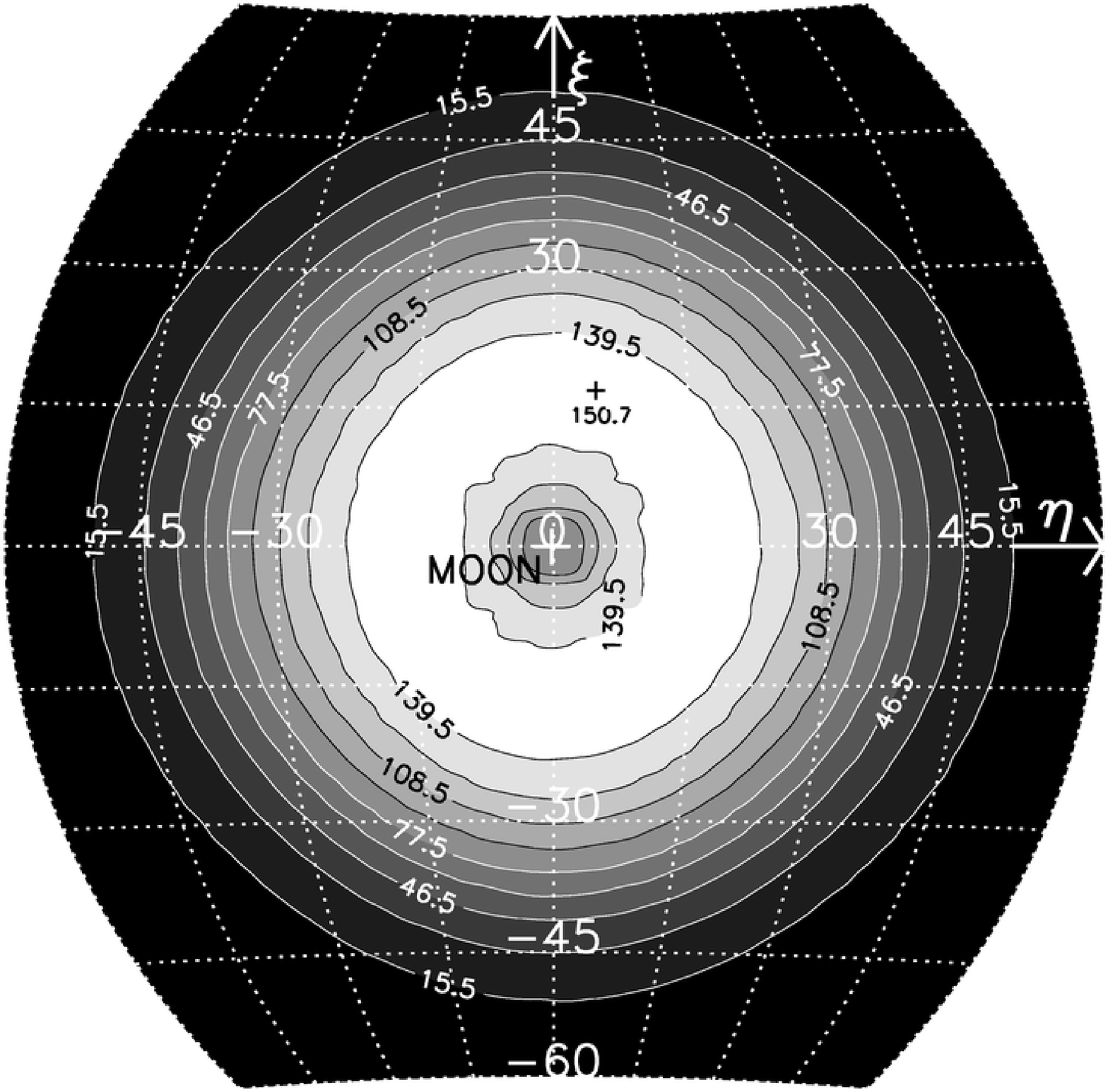}
\includegraphics[clip=true,width=2.5in]{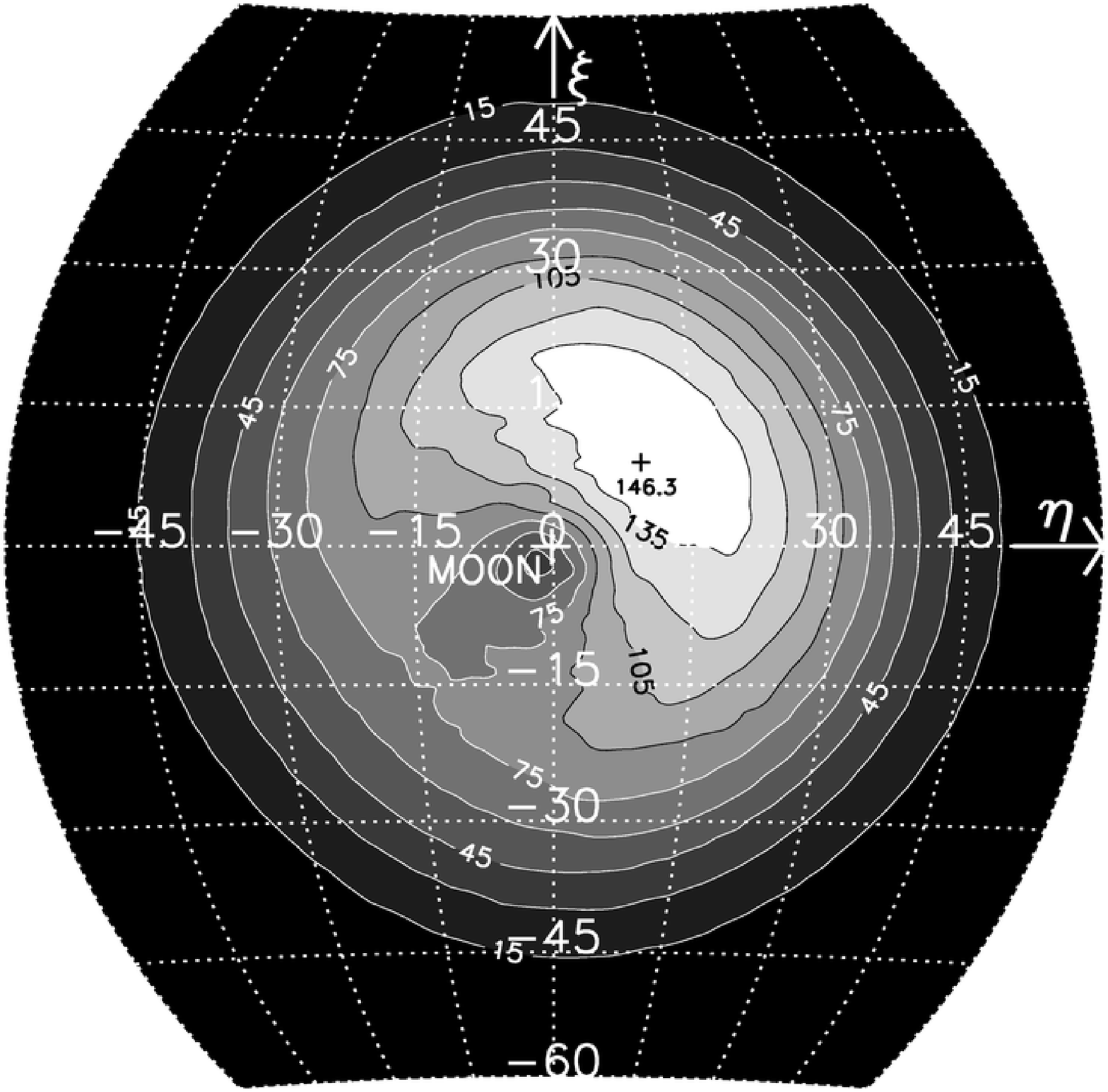}
 \caption{Effective area (km$^2$) of the experiment to $10^{22}$~eV neutrinos as a function of particle arrival direction, for (top) February, and (bottom) May. The telescope beams are centred on the Moon's centre $(0,0)$ and upper-right limb ($0.17,0.17$) respectively.}
 \label{fig0405}
\end{figure}

Since we were targeting particular sources, the instantaneous `beam-shape' (effective area as a function of particle arrival direction) of the ATCA-Moon system was also of interest. Again using the methods of Ref.\ \cite{JP}, we calculated this function for our observations, and plot it in Fig.\ \ref{fig0405}.
The limitation in targeting Cen A comes from the closest approach of the Moon being of order $30^{\circ}$, where our sensitivity falls significantly, while for the Galactic Centre region, it is due to the Moon passing too close to Sgr A. Nonetheless, our exposure to these regions was significantly enhanced, as is shown in the next section.

\section{Limit on an Isotropic Flux}

\begin{figure}[!ht!b]
 \centering
 \includegraphics[width=2.5in]{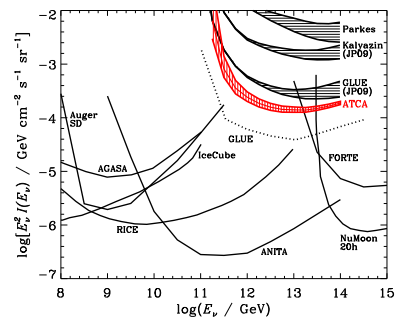}
 \caption{Isotropic limits on an UHE $\nu$ (see text).}
 \label{isolimatca}
\end{figure}

Subtracting the trigger dead-time from the raw observation times allowed us to calculate a `model independent' ($2.3 (A_{\rm eff} t_{obs})^{-1}$) limit on an isotropic flux of UHE $\nu$ based on our null result. This is compared to existing limits in Fig.\ \ref{isolimatca}. The range for the ATCA observations reflects minimum and maximum sensitivities; for the GLUE \cite{GLUE}, Parkes \cite{Parkes}, and Kalyazin \cite{Kalyazin} experiments (estimates from \cite{JP}), it reflects the inclusion/exclusion of a sub-regolith layer. While these limits exceed those of previous lunar Cherenkov experiments in the $E_{\nu} \lesssim 3~10^{22}$~eV energy range (where NuMoon has stronger limits \cite{NuMoon}), a $100$-fold increase in observation time (approximately a full year's observation) would be needed at the current sensitivity to compete with the limits from ANITA \cite{ANITA}.

\begin{figure*}[!ht!b]
 \centering
 \includegraphics[width=5in, clip=true]{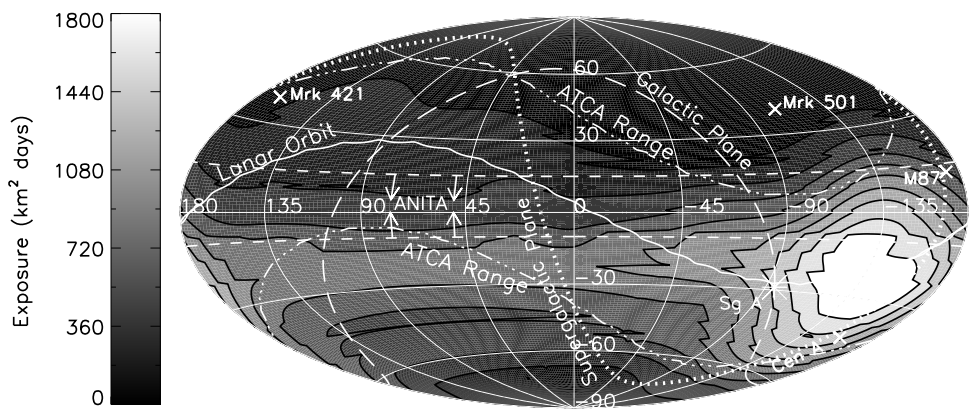}
 \caption{Combined exposure (km$^2$-days) from our LUNASKA ATCA observations, RICE \cite{RICE}, Parkes \cite{Parkes}, and GLUE \cite{GLUE} using our simulation results \cite{JP}, to $10^{23}$~eV neutrinos.}
 \label{directional_exposure}
\end{figure*}

While our isotropic limits are not competitive, the observation times and pointing positions were designed to target Cen A and the broad Sgr A region, as mentioned above. Integrating the instantaneous sensitivity of Fig.\ \ref{fig0405}
 with time -- given the known lunar position and orientation of the antenna beams -- allows the total exposure (km$^2$-days) as a function of celestial coordinates to be determined. This is plotted for neutrinos of $10^{23}$~eV in Fig.\ \ref{directional_exposure}, with approximate results for Parkes, GLUE, and RICE. The approximate declination range of ANITA is also shown, where the limit will dominate \cite{ANITA}. Table \ref{ATCA_exposures_table} compares the exposure from these observations to Cen A and Sgr A with that from RICE \cite{RICE}, and our estimates for GLUE. Evidently, we have been successful in increasing our relative exposure to these potential sources of UHE particles.

\begin{table*}
\begin{center}
\begin{tabular}{c | c c c c | c c c c}
\hline
\multirow{2}{*}{$E_{\nu}$} & \multicolumn{4}{c |}{Sgr A} & \multicolumn{4}{c}{Cen A} \\
		& GLUE	& RICE	& ATCA	& Total & GLUE	& RICE	& ATCA	& Total \\
\hline \hline
$10^{21}$~eV	& 0.5	& 195	& 2.9	& { 198}	& 0.015	& 242	& 6.9	& { 249}	\\
$10^{22}$~eV	& 14	& 333	& 54	& { 401}	& 2.1	& 242	& 111	& { 355}	\\
$10^{23}$~eV	& 175	& 706	& 409	& { 1290}	& 43	& 512	& 745	& { 1300}	\\
\hline
\end{tabular}
\end{center}
\caption[Experimental exposures to UHE neutrinos from Cen A and Sgr A.]{Experimental exposures (km$^2$-days) of GLUE, RICE, and the LUNASKA ATCA observations to UHE neutrinos at discrete energies from Cen A and Sgr A.}
\label{ATCA_exposures_table}
\end{table*}

\section{Discussion and Conclusion}
We have successfully used an array of small, wide-bandwidth antennas to search for UHE particles using the lunar Cherenkov technique. 
That the limits on an isotropic flux are not competitive is expected, since these were only trial observations. However, the required observation time to set a competitive limit with the current system is probably impractical at this stage. We therefore concentrate our efforts on improving our sensitivity before conducting observations of serious length.

\section*{Acknowledgments}
The Australia Telescope Compact Array is part of the Australia Telescope which is funded by the Commonwealth of Australia for operation as a National Facility managed by CSIRO. This research was supported by the Australian Research Council's Discovery Project funding scheme (project
numbers DP0559991 and DP0881006). J.A-M thanks Xunta de Galicia (PGIDIT 06 PXIB 206184 PR) and
Conseller\'\i a de Educaci\'on (Grupos de Referencia Competitivos -- Consolider Xunta de Galicia 2006/51).

\end{document}